\documentclass[aps,prl,reprint]{revtex4-1}
\usepackage{mathtools}

\begin{document}
\title{Landauer conductance in the complex domain: \\A path to find closed-form solutions}
\author{Mauricio J. Rodr\'iguez}
\author{Bryan D. Gomez}
\author{Carlos Ram\'irez}\email{carlos@ciencias.unam.mx}
\affiliation{Departamento de F\'isica, Facultad de Ciencias, Universidad Nacional Aut\'onoma de M\'exico, Apartado Postal 70542, 04510 Ciudad de M\'exico, M\'exico}

\begin{abstract}
The Landauer formula allows us to describe theoretically the conductance in terms of the transmission function in a mesoscopic system. We propose a general method to evaluate the transmission function in the complex domain for systems connected to semi-infinite atomic chains. This reveals the presence of complex-conjugated pairs of simple poles that are responsible for transmission peaks in the real-domain evaluations. This leads us to formulate a closed-form expression for the transmission function.
\end{abstract}

%Use showkeys class option if keyword
                              %display desired
\maketitle

%\section{\label{sec:level1}Introduction}
In recent decades, advances in miniaturization techniques down to the atomic level at low temperature have driven us to the mesoscopic scale, where quantum coherence is maintained, thus making it necessary to understand the effects caused by the emergence of complex interference phenomena.

Electronic transport in mesoscopic systems is theoretically addressed in state of the art calculations \cite{Li2019,wu2020,Li2020,Borsoi2020} via the famous Landauer formula for the conductance $(G)$ \cite{Datta2005,Ryndyk2016},
\begin{equation}
    G=G_0\int_{-\infty}^{\infty} T(E)\left(-\frac{\partial f}{\partial E}\right)dE,
    \label{landauer1}
\end{equation}
where $G_0=\frac{2e^2}{h}$ is the quantum of conductance, $f(E,\theta,\mu)$ is the Fermi distribution at energy $E$, temperature $\theta$ and chemical potential $\mu$, while $T(E)$ is the transmission function, that can be straightforwardly calculated from the scattering matrix (S-matrix) of the system \cite{Datta2005,Ryndyk2016,Lewenkopf2013}, even though it is frequently computed by using the non-equilibrium Green function \cite{Lewenkopf2013,Kazymyrenko2008,Thorgilsson2014}. For the case of zero temperature, the Fermi distribution becomes a step function, and then the Landauer conductance becomes \cite{Imry1999,Datta2005,Ryndyk2016}
\begin{equation}
    G=G_0T(E_F),
    \label{landauer2}
\end{equation}
where $E_F$ is the Fermi energy. Shot noise \cite{Blanter2000,Lumbroso2018}, Fano factor\cite{Lewenkopf2013}, band structure\cite{Ramirez2018}, total density of states\cite{Lewenkopf2013,Buttiker1999}, and other physical quantities relevant to characterize electron transport\cite{Wigner1955,Gopar1996,Buttiker1993,Texier2016} can also be obtained from the transmission function or from elements of the S-matrix of the system. Therefore, having access to the analytical expression of $T(E)$ would be ideal to enhance our comprehension of transport phenomena. Due to very cumbersome algebra, this has been done only in some systems \cite{Macia1996,Choi1999,Onipko2000,Chakrabarti2003,Yin2010,Jana2011,Sajjad2013,Ramirez2013,Sanchez2014,Wang2015,Deng2015,Munoz2017,OjedaSilva2018,Mrabti2018,Amini2019,Tamura2019,Mrabti2020}.

Recently, by using the recursive S-matrix method (RSMM)\cite{Ramirez2017,Ramirez2018}, we have proposed a method to find the exact Taylor expansion of $T(E)$ to arbitrary order\cite{Ramirez2020}, which, via a Sommerfeld expansion\cite{Nolting2018}, is useful to express the solution to Eq. (\ref{landauer1}) at the low temperatures limit as
\begin{equation}
    \frac{G}{G_0}=T(\mu)+\frac{\pi^2}{6}(k_B \theta)^2T''(\mu)+\frac{7\pi^4}{360}(k_B \theta)^4T''''(\mu)+...
    \label{landauer3}
\end{equation}
The analysis of Taylor expansions reveal a finite convergence radius $(R)$, giving insights of singularities in the complex domain of the transmission function \cite{Ramirez2020}. Consequently, Eq. (\ref{landauer3}) is only useful if the thermal energy $(k_B\theta)$ is small in comparison to $R$.

In this letter, we start with a brief review of the RSMM and determine how to use it to obtain the transmission function in the complex-domain. Then, we determine the transmission spectrum of tight-binding chains with hopping integrals following periodic and aperiodic sequences, revealing the presence of singularities. Finally, based on the behaviour of singularities, we propose and validate a closed form expression for the transmission function.

The S-matrix of a two terminal system can be written as
\begin{equation}
    S(E)=\begin{pmatrix} \mathbf{r}(E) & \mathbf{t}'(E) \\ \mathbf{t}(E) & \mathbf{r}'(E) \end{pmatrix}
\end{equation}
where $\mathbf{r}$ ($\mathbf{r}'$) is the reflection matrix and $\mathbf{t}$ ($\mathbf{t}$') is the transmission matrix at energy $E$ for incident wave from terminal 1 (2). In terms of the elements of the S-matrix, the transmission function is \cite{Ryndyk2016}
\begin{equation}\label{Transmision}
  T(E)=Tr[\mathbf{t}^{\dagger}(E)\mathbf{t}(E)]=\sum_{n=1}^{C_2} \sum_{m=1}^{C_1} |t_{n,m} (E)|^2,
\end{equation}
where $C_1$ ($C_2$) is the number of open conduction channels in terminal 1 (2).

%\subsection{\label{RSMM}Recursive S-matrix method (RSMM)}
\begin{figure}[tp]
    \center
    \includegraphics[width=0.45\textwidth]{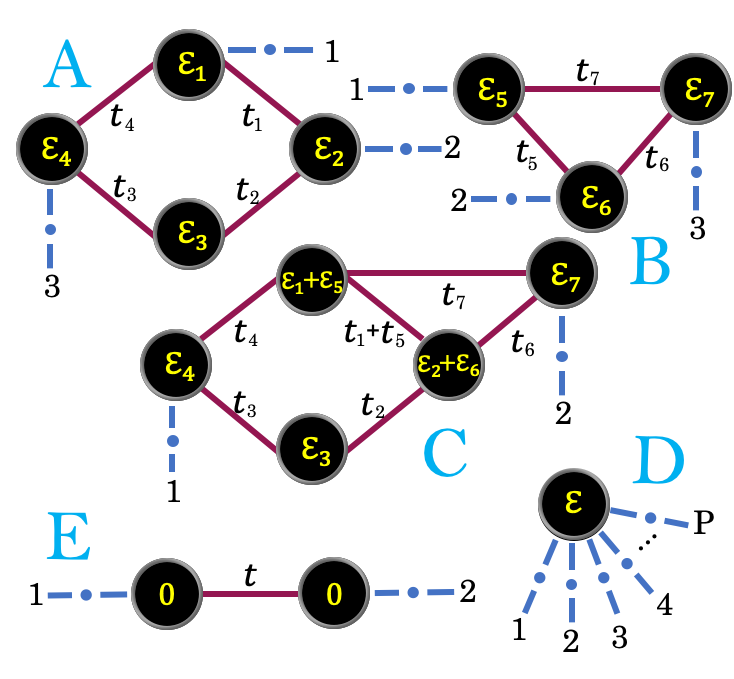}
    \caption{Examples of tight-binding structures connected to semi-infinite atomic chains with null site-energies and hopping integrals $t_C$ (blue dashed-dot-dashed lines). Incomming and outgoing waves in these systems travel along these coupled chains, and their coefficientes are related by the S-matrix of each system. RSMM consists in equaling the coefficients of the first $N$ outgoing waves of structure $A$ to those of the first $N$ incoming waves of structure $B$, and vice-versa. In this example, the S-matrix of structure $C$ is obtained by using the RSMM on structures $A$ and $B$ with $N=2$. Structures D and E represent respectively the site and the bond structures.}
    \label{fig1}
\end{figure}

The RSMM allows us to compute the S-matrix of tight-binding systems from the S-matrices of its subsystems \cite{Ramirez2017}. Figure \ref{fig1} exemplifies the RSMM, where structures $A$ and $B$ are glued together to obtain structure $C$. Mathematically, if S-matrices of structures $A$ and $B$ are respectively given by
\begin{equation}
    \mathbf{S}^{A}=\begin{pmatrix}\mathbf{S}^{A}_{11} & \mathbf{S}^{A}_{12}\\\mathbf{S}^{A}_{21} & \mathbf{S}^{A}_{22} \end{pmatrix} \mbox{ and } \mathbf{S}^{B}=\begin{pmatrix} \mathbf{S}^{B}_{11} & \mathbf{S}^{B}_{12} \\ \mathbf{S}^{B}_{21} & \mathbf{S}^{B}_{22} \end{pmatrix},
\end{equation}
where sub-matrices $\mathbf{S}^{A/B}_{11}$ are $N\times N$ matrices, then S-matrix of structure $C$ becomes 
\begin{equation}
    \mathbf{S}^{C}=\begin{pmatrix} \mathbf{S}^{C}_{11} & \mathbf{S}^{C}_{12} \\ \mathbf{S}^{C}_{21} & \mathbf{S}^{C}_{22} \end{pmatrix},
\end{equation}
with
\begin{equation}
    \begin{array}{l} \mathbf{S}^{C}_{11}=\mathbf{S}^{A}_{22}+\mathbf{S}^{A}_{21}(\mathbf{I}-\mathbf{S}^{B}_{11}\mathbf{S}^{A}_{11})^{-1}\mathbf{S}^{B}_{11}\mathbf{S}^{A}_{12},\\\mathbf{S}^{C}_{12}=\mathbf{S}^{A}_{21}(\mathbf{I}-\mathbf{S}^{B}_{11}\mathbf{S}^{A}_{11})^{-1}\mathbf{S}^{B}_{12},\\\mathbf{S}^{C}_{21}=\mathbf{S}^{B}_{21}(\mathbf{I}-\mathbf{S}^{A}_{11}\mathbf{S}^{B}_{11})^{-1}\mathbf{S}^{A}_{12},\\\mathbf{S}^{C}_{22}=\mathbf{S}^{B}_{22}+\mathbf{S}^{B}_{21}(\mathbf{I}-\mathbf{S}^{A}_{11}\mathbf{S}^{B}_{11})^{-1}\mathbf{S}^{A}_{11}\mathbf{S}^{B}_{12}.\\ \end{array}
\end{equation}
Any tight binding structure, including multiterminal systems with general leads \cite{Ramirez2018}, can be modeled by starting from the site and bond structures \cite{Ramirez2017}, represented respectively by structures D and E of Fig. \ref{fig1}, and whose S-matrices are analytically given  by
\begin{equation}\label{Ssite}
    \mathbf{S}_{n,m}^\textnormal{site}=\frac{t_C\left(e^{i\kappa}-e^{-i\kappa}\right)}{\varepsilon-E+Pt_Ce^{i\kappa}}-\delta_{n,m},
\end{equation}
and
\begin{equation}\label{Sbond}
    \mathbf{S}_{n,m}^\textnormal{bond}=\begin{pmatrix}r&\frac{t}{t_c}\left(e^{i\kappa}+re^{-i\kappa}\right) \\ \frac{t}{t_c}\left(e^{i\kappa}+re^{-i\kappa}\right)&r
    \end{pmatrix},
\end{equation}
where $r=-(t^2-t_c^2)/(t^2-t_c^2e^{-2i\kappa})$ and $E=2t_c\cos \kappa$. For Taylor expansions and the extension to the complex domain discussed below, it is convenient to use that
\begin{equation}
    e^{i\kappa}=\frac{E}{2t_c}+i\sqrt{1-\left(\frac{E}{2t_c}\right)^2}.
\end{equation}

%\subsection{Taylor series of $T(E)$}
Recently, we extended the RSMM to find the order-$M$ Taylor expansion of the S-matrix about $E_0$ \cite{Ramirez2020},
\begin{equation}
    \mathbf{S}(E;E_0)=\sum_{j=0}^M \mathbf{S}_j^{(E_0)}(E-E_0)^j.
\end{equation}
Consequently, each component of the transmission matrix can be expressed as a Taylor series,
\begin{equation}
    t_{n,m}(E;E_0)=\sum_{j=0}^M t_{n,m,j}^{(E_0)}(E-E_0)^j.
    \label{transreal}
\end{equation}
If $E$ and $E_0$ are real numbers,
\begin{equation}
    t_{n,m}^*(E;E_0)=\sum_{j=0}^M \left(t_{n,m,j}^{(E_0)}\right)^*(E-E_0)^j.
    \label{transcomplex}
\end{equation}
Hence, by using Eqs. (\ref{transreal}) and (\ref{transcomplex}) in (\ref{Transmision}), the Taylor expansion of the transmission function becomes
\begin{equation}\label{transTaylorreal}
  T(E;E_0)= \sum_{j=0}^M \sum_{k=0}^j \sum_{n=1}^{C_2} \sum_{m=1}^{C_1}  t_{n,m,k}^{(E_0)} \left(t_{n,m,j-k}^{(E_0)}\right)^* (E-E_0)^j.
\end{equation}

%\subsection{\label{extensionC}Extension to $\mathbb{C}$-domain}
From Eq. (\ref{transTaylorreal}) the analytical extension of the transmission function to the complex domain becomes
\begin{equation}
  T(Z;Z_0)= \sum_{j=0}^M \sum_{k=0}^j \sum_{n=1}^{C_2} \sum_{m=1}^{C_1}  t_{n,m,k}^{(Z_0)} \left(t_{n,m,j-k}^{(Z_0^*)}\right)^* (Z-Z_0)^j,
  \label{TaylorComplex}
\end{equation}
where $Z$ and $Z_0$ are complex numbers. In particular, notice that
\begin{equation}
  T(Z;Z)=\left[T(Z^*;Z^*)\right]^*,
  \label{Tconj}
\end{equation}
\textit{i.e.}, transmission function evaluated at $Z$ is the complex conjugate of the transmission function evaluated at $Z^*$.

\begin{figure*}[t]
    \center
    \includegraphics[width=1.\textwidth]{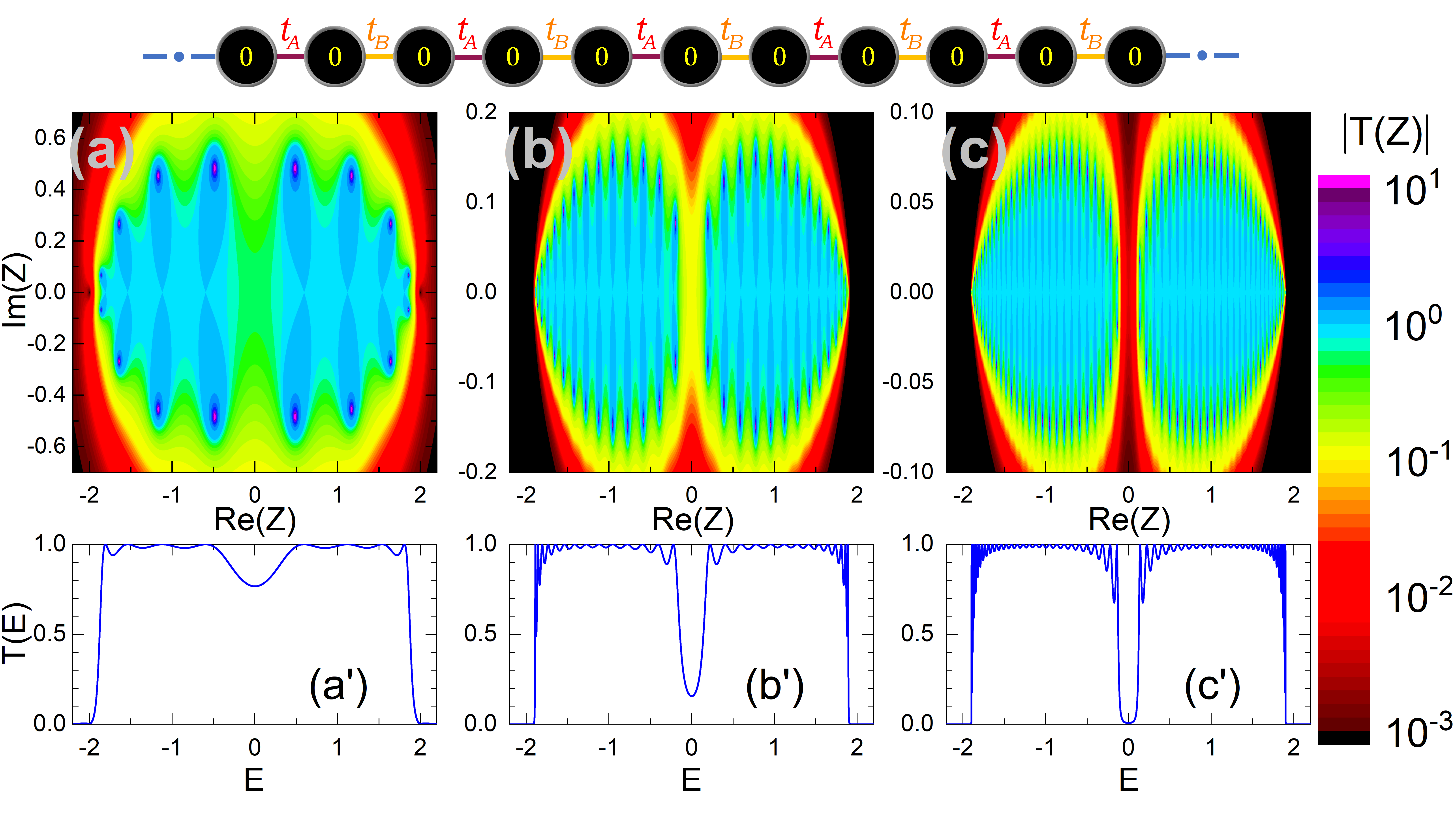}
    \caption{Transmission function in (a,b,c) the complex and (a',b',c') the real domains of atomic chains with hopping integrals following the periodic sequence for generations (a,a') 5, (b,b') 15 and (c,c') 30. (On top) The tight-binding structure for the case of generation 5. $E$ and $Z$ are given in units of $t_A$}
    \label{fig2}
\end{figure*}

\begin{figure*}[t]
    \center
    \includegraphics[width=1.\textwidth]{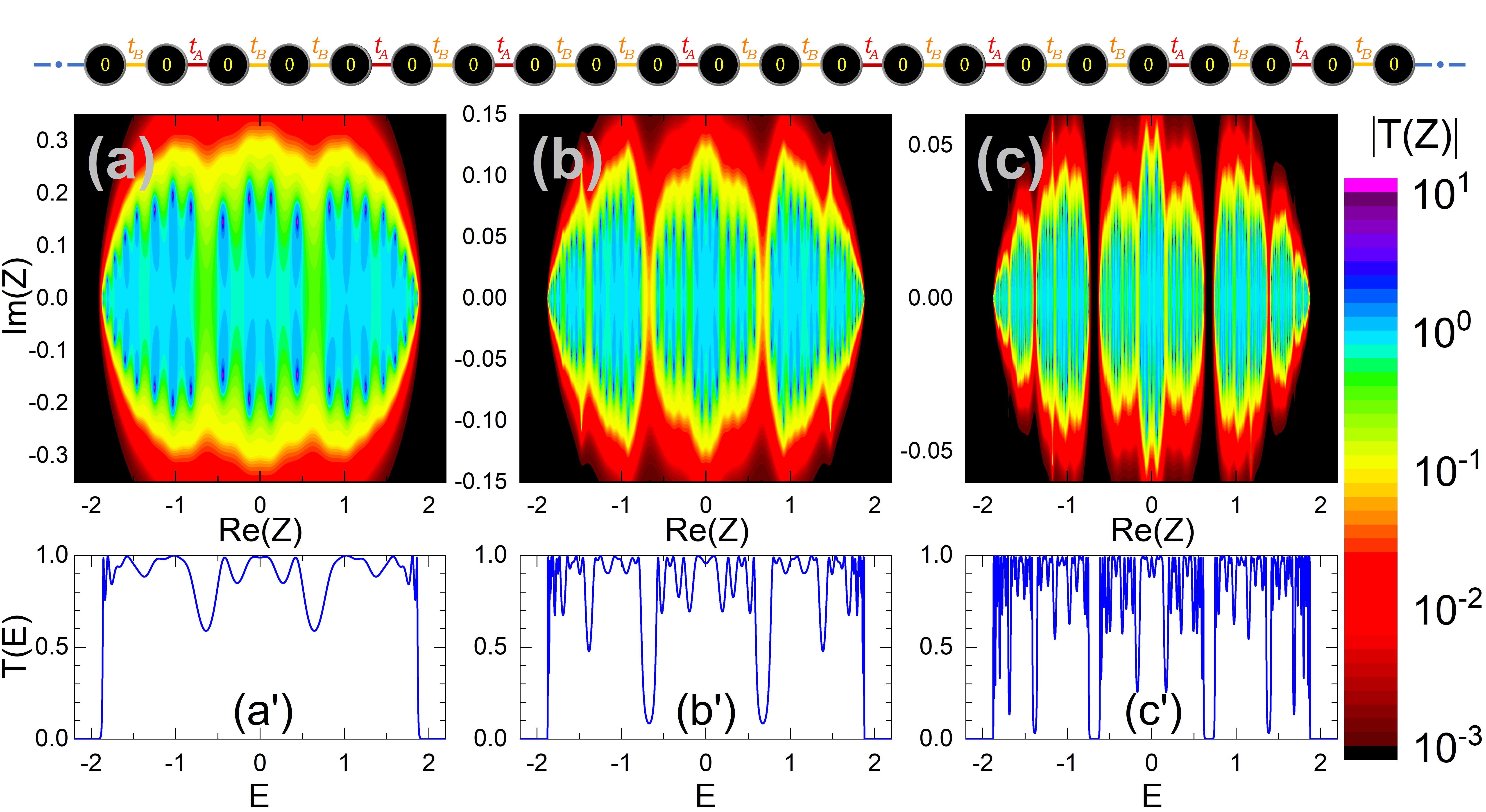}
    \caption{Transmission function in (a,b,c) the complex and (a',b',c') the real domains of atomic chains with hopping integrals following the Fibonacci sequence for generations (a,a') 8, (b,b') 10 and (c,c') 12. (On top) The tight-binding structure for the case of generation 8. $E$ and $Z$ are given in units of $t_A$}
    \label{fig3}
\end{figure*}

In the following, we present the transmission function in the complex domain for atomic chains with null site energies, and two hopping integrals $t_A$ and $t_B=0.9t_A$ ordered in periodic and Fibonacci sequences, as illustrated in top of Figs. \ref{fig2} and \ref{fig3}. Leads (represented by blue dot-dashed lines) are semi-infinite atomic chains with null site energies and hopping integrals $t_C=t_A$.

%\subsection{Periodic Sequence}
The $j$-th generation of the periodic sequence $(S_{j})$, is obtained by the rule
\begin{equation}
    S_{j}=S_{1}^{j} ,\,\,\forall j\ge 1,
    \label{ordenadosecuencia}
\end{equation}
where $S_1=AB$ and $S_{1}^{j}$ means to repeat $j$-times the sequence $S_{1}$. For example, $S_{2}=S_{1}^{2}=ABAB$, $S_{3}=S_{1}^{3}=ABABAB$. Calculation of these structures can be optimized by using the doubling algorithm \cite{Rumpf2011}. Figure \ref{fig2} shows, in color scale, the absolute value of transmission function in the complex domain of chains with hopping integrals following the periodic sequence, for generations (a) 5, (b) 15 and (c) 30. Evaluations in the real domain are shown in (a'), (b') and (c'), respectively. Notice in all cases the mirror symmetry with respect to the real axis, as expected from Eq. (\ref{Tconj}). Results in the complex domain reveal multiple singularities. According to Eq. (\ref{Tconj}), they come in complex conjugated pairs. Observe that each pair is related to a peak of transmission in the real domain. These peaks are sharper for singularities closer to the real axis. For longer systems, the number of these singularities is greater and they are closer to the real axis, which correspond to an increase of oscillations in the real domain. Note that singularities are located following a band-like structure, with a gap around $\operatorname{Re}(Z)=0$ that is noticeable in the real domain when the chain is longer.

%\subsection{Fibonacci sequence}

The Fibonacci sequence is obtained from the concatenation rule
\begin{equation}
    S_{i}=S_{i-1}S_{i-2},\,\,\forall i\ge 3
     \label{fibonaccisecuencia}
\end{equation}
where $S_{1}=A$ and $S_{2}=B$. For example, $S_{3}=BA$, $S_{4}=BAB$, and $S_{5}=BABBA$. Figure \ref{fig3} shows, in color scale, the absolute value of transmission function in the complex domain of chains with hopping integrals following the Fibonacci sequence, for generations (a) 8, (b) 10 and (c) 12. Evaluations in the real domain are shown in (a'), (b') and (c'), respectively. As occurred with the periodic case, in all cases there is a mirror symmetry with respect to the real axis, and peaks of transmission in the real domain are related to the position of singularities in the complex domain, with sharper peaks for singularities closer to the real axis. On the other hand, the location of singularities this time follows a fractal behaviour, which is more evident in the complex than in the real domain. By comparing Figs. \ref{fig2}(c) and \ref{fig3}(c), we observe that for chains of similar length, singularities are closer to the real axis for the case of the Fibonacci sequence. As a consequence, oscillations in the real domain are smoother in the periodic sequence.
  
  It is worth to mention that in all cases, if the singularity has real part in $[-2t_C,2t_C]$ (where there are open channels in the leads) its imaginary part is always greater than zero, avoiding us to find singularities in conductance spectra. On the other hand, some systems have singularities with real part beyond $[-2t_C,2t_C]$, but always in the real domain. Please see the Supplemental Material where the location of the singularities for some systems is numerically given.
  
%\section{\label{sec:level4}Closed form expression of the transmission function}
\begin{figure}[htp]
    \centering
    \includegraphics[width=0.45\textwidth]{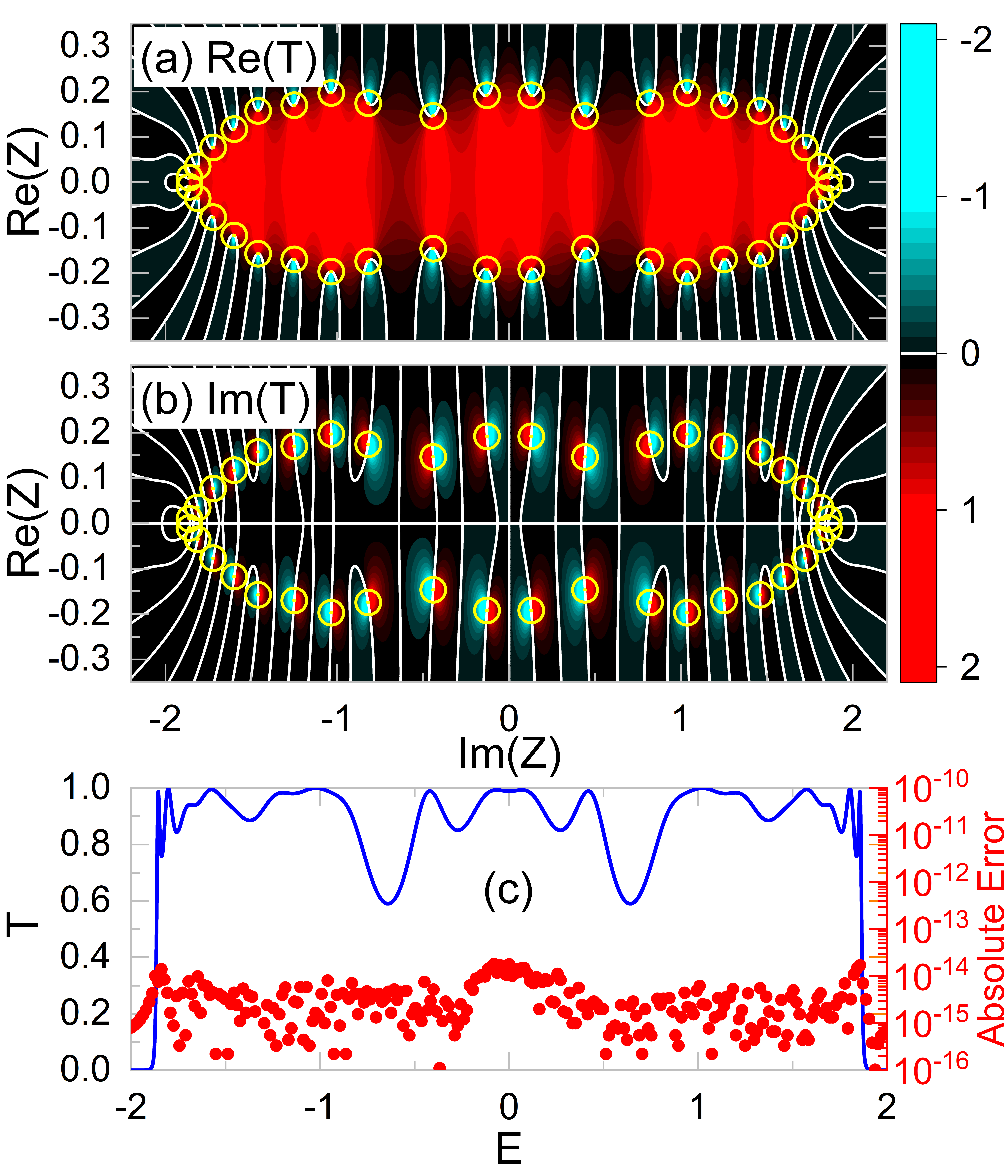}
    \caption{(a) Real and (b) imaginary parts of the transmission function, for the same case of Fig. \ref{fig3}(a), where open circles are centered at the position of the poles and nodal lines are shown in white color. (c) Transmission function (solid line) and absolute error (solid circles) obtained by using the closed-form expression in Eq. (\ref{closedform}).}
    \label{figure4}
\end{figure}
Figures 4(a) and 4(b) show respectively the real and the imaginary part of the transmission function in the complex domain for the same structure of Fig. \ref{fig3}(a), with nodal lines shown in white color. Open circles are centered at each singularity. Observe that each singularity is intersected by one nodal line of the real part and one of the imaginary part of the transmission function. This means that all singularities are simple poles. The same occurs in other chains. Consequently, we propose that the transmission function in atomic chains is generated by a linear combination of the effect caused by the poles, \textit{i.e.},

\begin{equation}
    T(Z)=A_0+\sum_{i=1}^p\frac{A_i}{Z-\omega_i}.
    \label{closedform}
\end{equation}
where $p$ is the number of poles, $\omega_i$ is the position of the $i$-th pole, and $\{A_i\}$ are scalars. From Eq. (\ref{closedform}),
\begin{equation}
    F(Z)\equiv\frac{1}{T(Z)}=\frac{Z-\omega_j}{A_j+A_0(Z-\omega_j)+\sum_{\substack{i=1\\(i\neq j)}}^p A_i\frac{Z-\omega_j}{Z-\omega_i}}.
    \label{invclosedform}
\end{equation}
with $j=1,2,...,p$. Observe that $F(\omega_j)=0$, \textit{i.e.}, the poles of T(Z) can be found by searching for the zeroes of $F(Z)$. Using Eq. \ref{TaylorComplex}, the exact first order Taylor series of $T(Z)$ can be computed. Then, by employing automatic differentiation \cite{Makino2003}, the exact value of F(Z) and its derivative is determined. These values can be used within the Newton-Raphson root-finding method to find the position of poles to any desired precision in a few iterations. On the other hand, notice that
\begin{equation}
\frac{dF(Z)}{dZ}=\frac{A_j+\sum_{\substack{i=1\\(i\neq j)}}^p A_i\left(\frac{Z-\omega_j}{Z-\omega_i}\right)^2}{\left[A_j+A_0(Z-\omega_j)+\sum_{\substack{i=1\\(i\neq j)}}^p A_i\frac{Z-\omega_j}{Z-\omega_i}\right]^2}.
\end{equation}
Therefore, for $j=1,2,...,p$,
\begin{equation}
    A_j=\left[\frac{dF(Z)}{dZ}\Bigr|_{Z=w_j}\right]^{-1},
\end{equation}
which, as explained above, can be exactly calculated. Finally, $A_0$ is obtained by comparing the summation in Eq. (\ref{closedform}) to the exact evaluation of $T(E_0)$ at some energy $E_0$.

Figure 4(c) shows the transmission function (solid line) obtained by using Eq. \ref{closedform} (values of $\omega_j$ and $A_j$ for this figure are given in the Supplemental Material), which is in full agreement to that shown in Fig. \ref{fig3}(a'). Observe that the absolute error (red dots) reach the machine double-precision, validating the closed-form proposal of Eq. \ref{closedform}. It is worth to mention that we have also validated this expression with other structures, some of them are presented in the Supplemental Material. Until now, we have not found a structure whose transmission is not reproduced by this closed-form expression.

%\section{\label{sec:conclusion}Conclusion}
In summary, we have extended the recursive scattering matrix method to the complex domain for arbitrary systems connected to semi-infinite atomic chains. Transmission spectra in the complex domain has simple poles in all systems. Poles form a band-like structure in periodic chains, and present a fractal behaviour for chains with hopping integrals following the Fibonacci sequence. Finally, we propose and validate numerically a closed-form expression for the transmission function. The number of parameters in this expression is $2p+1$ where $p$ is the number of poles. As in any analytical solution, this expression could improve our understanding about transport properties. For example, it could be useful to accurately and efficiently obtain the behaviour of conductance at different temperatures. We expect these results can be extended to systems with general leads, a study that is currently under development.

%\begin{acknowledgments}
This work was supported by UNAM-PAPIIT IN116819. Computations were performed at Miztli under project LANCAD-UNAM-DGTIC-329.

\bibliographystyle{apsrev4-1} % Tell bibtex which bibliography style to use
\bibliography{References} % Tell bibtex which .bib file to use (this one is some example file in TexLive's file tree)

\end{document}